\documentclass[twocolumn,prb]{revtex4}% Physical Review B

\usepackage{graphicx}% Include figure files
\usepackage{epsfig}
\usepackage{dcolumn}% Align table columns on decimal point
\begin{document}

%\preprint{APS/123-QED}

\title{Dephasing of conduction electrons by magnetic impurities in Cu/Ni and Cu/Cr samples:
Influence of spin-glass transition on the superconducting proximity effect}% Force line breaks with \\
\author{I. Sosnin}
\email{i.sosnin@rhul.ac.uk}
\author{P. Nugent}
\author{J. Zou}
\author{V.T. Petrashov}%
\affiliation{Physics Department, Royal Holloway University of
London, Egham, Surrey, TW20 0EX }
\author{A.F. Volkov}
\affiliation{Theoretische Physik III, Ruhr-Universit\"{a}t Bochum,
D-44780 Bochum, Germany}

\affiliation{Institute of Radioengineering and Electronics of the
Russian Academy of Sciences, 125009 Moscow, Russia }

\date{\today}% It is always \today, today,

\begin{abstract}
The dependence of the superconducting proximity effect on the
amount of magnetic impurities in the normal part of Andreev
interferometers has been studied experimentally. The dephasing
rates obtained from fitting experimental data to quasiclassical
theory of the proximity effect are consistent with the spin flip
scattering from Cr impurities forming a local moment in the Cu
host. In contrast, Ni impurities do not form a local moment in Cu
and as a result there is no extra dephasing from Ni as long as
Cu/Ni alloy remain paramagnetic.
\end{abstract}

\maketitle
\section{INTRODUCTION}
Electron dephasing has been one of the most important problems of
mesoscopic physics since its emergence in the 1980s. The main
sources of dephasing have been identified as inelastic scattering
due to electron-electron and electron-phonon interactions and
scattering by magnetic impurities \cite{1}. These have been
carefully studied experimentally using the weak localization
correction to the conductance of mesoscopic structures \cite{2}.
The topic has received renewed interest since the proposal of
quantum computing (see e.g. \cite{3}). Dephasing is one of the
major obstacles in building a working solid-state quantum bit.
Practically, the phase breaking time, $\tau_{\phi}$, is often
limited by the presence of even tiny amounts of magnetic
impurities \cite{4, 5}.

The mechanism of dephasing by magnetic impurities has been studied
extensively using weak localization \cite{6} and the suppression
of the superconducting critical temperature \cite{7}. The
dephasing rate in these experiments has been identified with the
spin-flip rate obtained from the low temperature logarithmic
increase in resistivity, the Kondo effect \cite{8}. Theoretically
the problem of conductance in mesoscopic systems with magnetic
impurities has been studied in various ranges of temperature and
impurity concentration \cite{9, 10, 11, 12}. Recently, the effect
of Kondo impurities on the superconducting proximity effect has
been addressed qualitatively in the Au/Fe system \cite{13}.

Here we present an experimental study of the effect of the
magnetic impurities on the coherent part of the conductance of a
normal metal in proximity to a superconductor. We have
investigated a wide range of Kondo temperatures, $T_{K}$, for the
two chosen systems: $T_{K}\approx$10K for Cu/Cr and
$T_{K}\approx$1000K for Cu/Ni \cite{14}. The dephasing rate
obtained by fitting amplitude of resistance oscillations with
magnetic field to the quasiclassical theory of the proximity
effect (see \cite{15} for a review on the superconducting
proximity effect) in both cases is compared to the spin-flip rates
estimated from Kondo effect.

\section{SAMPLE FABRICATION}

\begin{figure}
\centerline{\psfig{file=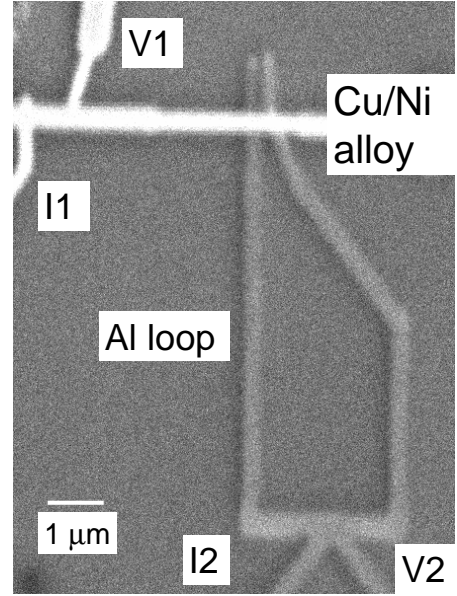,height=3.5 in}} \caption{SEM
micrograph of a measured sample. Current contacts are labeled
$I_{1}$ and $I_{2}$; voltage contacts, $V_{1}$ and $V_{2}$. The
area of superconducting loop is $12 \mu m^{2}$.} \label{Fig.1}
\end{figure}

The samples were fabricated using e-beam lithography and standard
processing. The geometry of the structures is shown in Fig. 1.
$I_{1}$ and $I_{2}$ are current leads; $V_{1}$ and $V_{2}$,
voltage leads. The distance between the N/S contacts is about
150nm, the length of normal wire between the voltage probes is
$L=2.6 \mu m$. The area of superconducting loop is about $12 \mu
m^{2}$. The first layer was the normal part, 40 nm thick Cu/Ni or
Cu/Cr alloy. The alloy films of various concentrations were
fabricated by simultaneous evaporation of Ni or Cr and Cu at fixed
rates to obtain the required concentration. To obtain clean
interfaces between the layers, the contact area was Ar$^{+}$
plasma etched before the deposition of the second
(superconducting) layer which was 60 nm thick Al film. In case of
Cu/Ni samples the resulting composition of the film was measured
using X-ray spectroscopy in an SEM with an accuracy better than
0.5\%. Cr concentration was determined by the slope of logarithmic
divergence of resistivity at low temperatures due to the Kondo
effect. The homogeneity of the alloy films produced by
co-evaporation has been checked by X-ray microanalysis and by
dc-extraction magnetometry on ferromagnetic alloys, and was found
to be better than 2\% (see also \cite{16}).

Evaporation sources for alloy film deposition were Cu wire of
99.996$\%$ purity and Ni wire of 99.98$\%$ purity from Advent Ltd.
Using typical analysis of the source purity provided by the
supplier we estimate that up to 10ppm of other magnetic impurities
such as Mn, Cr, or Fe can be introduced into the alloy film during
fabrication. Cr source was 99.98$\%$ pure from Testbourne Ltd.

\section{EXPERIMENTAL DATA}

Two sets of samples have been investigated: 4 Cu/Ni samples of
geometry shown in Fig. 1 with Ni concentrations 0, 4.0, 7.8, and
8.4 atomic percent; and 4 Cu/Cr samples of similar geometry with
Cr concentrations 0, 14, 18, and 41 ppm. All samples showed
magnetoresistance oscillations due to the superconducting
proximity effect. Figure 2 shows the reduced amplitude of
oscillations, $\Delta R/R_{N}$, for two samples with highest Ni
concentration ($R_{N}$ is the resistance of $N$ wire). The period
of oscillations corresponds to the magnetic flux quantum
$\Phi_{0}=hc/2e$ through the area of the superconducting loop.

\begin{figure}
\centerline{\psfig{file=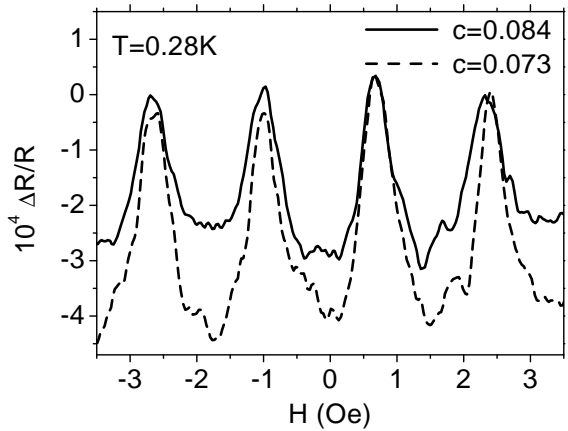,height=2.5 in}}
\caption{Magnetoresistance oscillations for two Cu/Ni samples at
$T$=0.28K.} \label{Fig.2}
\end{figure}

Figure 3 shows the attenuation of reduced amplitude of the
proximity effect (in logarithmic scale) with increasing
concentration of Ni impurities. Adding up to 8.4at\% of Ni to Cu
suppresses $\Delta R/R_{N}$ by about 100 times. Dots represent
experimental points, while the lines are theoretical fits which
will be described later. We estimate the relative inaccuracy in
this data to be at least 50$\%$ due to uncertainties in sample
geometry and the quality of interfaces.

\begin{figure}
\centerline{\psfig{file=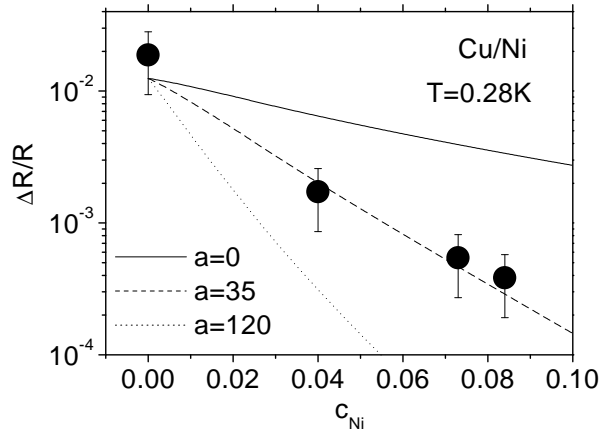,height=2.5in}} \caption{Dots:
Reduced amplitude of resistance oscillations in logarithmic scale
as function of impurity concentration at $T=0.28K$ for Cu/Ni
samples. Lines: theoretical fits using $\L_{\phi 0}=1.9\mu$m and
parameter $a$=0 (solid), 35 (dashed), and 120 (dotted).}
\label{Fig.3}
\end{figure}

\begin{figure}
\centerline{\psfig{file=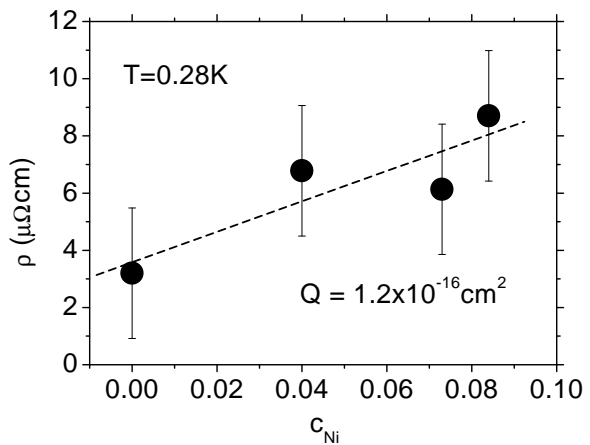,height=2.5 in}} \caption{Dots:
Low-temperature residual resistivity versus Ni impurity
concentration. Dashed line: best linear fit.} \label{Fig.4}
\end{figure}
To quantify the effect of Ni impurities on elastic scattering
times of conduction electrons we plot the value of residual
resistivity at low temperature versus impurity concentration, see
Fig. 4. The scattering of experimental values is due to the random
amount of disorder introduced during fabrication of thin films.
Based on statistics of resistivity measurements we estimate
relative inaccuracy of these values to be about 50$\%$. The data
allows us to calculate the elastic electron scattering cross
section, $Q_{e}$, for Ni impurities using the following relation
\cite{17}

\begin {equation}
Q_{e} = \frac{e^{2}}{mv_{F}}\frac{\partial \rho}{\partial c},
\end{equation}

where $e$ is electron charge, $m$ electron mass, $v_{F}$ is the
Fermi velocity, and $c = n_{imp}/n_{Cu}$ is the atomic
concentration of impurities ($n_{imp}$ and $n_{Cu}$ are
concentrations of impurities, Ni or Cr, and Cu atoms
respectively). The data in Fig. 5 gives the experimental slope
$\partial \rho/\partial c=0.7\mu \Omega cm / at \%$ in reasonable
agreement with $1.3\mu \Omega cm$ per at \% of Ni in Cu reported
previously \cite{18}. Substituting this value and $v_{F}=1.6\times
10^{8}cm/s$ \cite{19} into (3) we calculate
$Q_{e}=1.2\times10^{-16}cm^{2}$. This allows us to calculate the
elastic electron scattering rate on Ni atoms as
$\tau_{e}^{-1}=v_{F}n_{Cu}Q_{e}c$, using $n_{Cu}=8.45\times
10^{22}cm^{-3}$ (Ref. [19]).

\begin{figure}
\centerline{\psfig{file=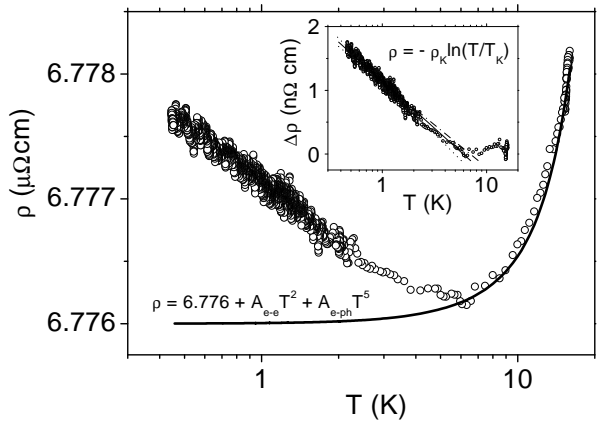,height=2.5 in}} \caption{Circles:
Resistivity of Cu+4$\%$Ni sample versus temperature in log scale.
Solid line: fit of high-temperature part of resistivity to
residual, electron-electron and electron-phonon contributions.
Inset: Kondo contribution to resistivity after subtraction of
residual, electron-electron and electron-phonon contributions.
Solid line: best fit to (2) with $\rho_{K}=0.67$n$\Omega$cm and
$T_{K}$=6.2K; Dotted line: $T_{K}$=5K; Dash-dotted line:
$T_{K}$=7K.} \label{Fig.5}
\end{figure}

A small amount of magnetic impurities forming a local moment in
the Cu host have been detected by the Kondo effect. The
temperature dependence of resistivity in the Cu/Ni alloy wires was
measured on separate long lines without any contacts with
superconductors made on the same chip simultaneously with the
Andreev interferometers. They showed a minimum at temperatures
between 3K and 10K for different concentrations. The position of
the minimum corresponds to the Kondo effect due to Fe or Cr
impurities. Since Cr has a smaller $T_{K}$ than Fe does, it will
contribute stronger to electron dephasing. Figure 5 shows
resistivity versus temperature graph for Cu+4$\%$Ni sample. Solid
line approximates high-temperature part of of resistivity to $\rho
= \rho_{0} + A_{ee}T^{2} + A_{eph}T^{5}$, where $\rho_{0}$=6.776
$\mu\Omega$cm is the residual resistivity. After subtracting this
fit from the data we fit the remaining dependence to the Kondo
formula \cite{8}

\begin {equation}
\Delta\rho=-\rho_{K}ln(T/T_{K}),
\end{equation}

with two fitting parameters, $\rho_{K}$ and $T_{K}$. The inset of
Fig. 5 shows the sensitivity of the fit to the values of the Kondo
temperature. Our data allows us to estimate $T_{K}$ within 1K
accuracy. From the slope of the resistivity versus log$T$ plot we
can calculate the concentration of Kondo impurities. For Cr in Cu
host the known value is 0.4$\pm$0.1 $n\Omega cm$ per decade of
temperature change per ppm \cite{20}. Figure 6 shows low
temperature resistivity increase due to Kondo effect after
subtracting electron-electron and electron-phonon contributions
for 4 different samples: pure Cu, Cu with 4$\%$ Ni and two Cu/Cr
ones. It follows from Fig. 6 that sample with 4$\%$ Ni can contain
up to 4ppm of Cr. Note also similar values of $T_{K}$ observed in
Cu/Ni and Cu/Cr samples. The effect of impurity-impurity
interaction is shown in Fig. 6 as deviation at low temperatures
from the behavior predicted by (2) for sample with 42ppm of Cr.
This may explain the decrease in dephasing rate for sample with 42
ppm of Cr (see Fig. 7). The pure Cu sample did not show a
resistivity minimum which means $c_{Cr}<1$ ppm in our case.

\begin{figure}
\centerline{\psfig{file=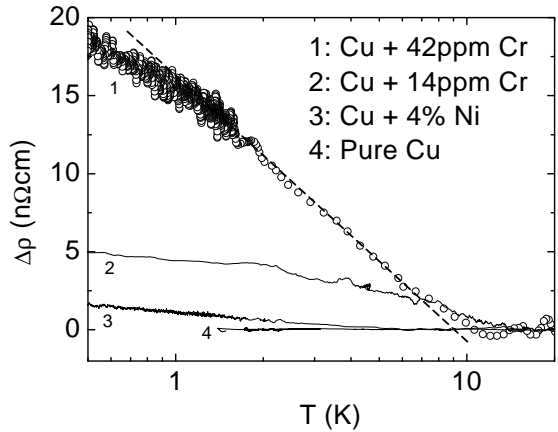,height=2.5 in}}
\caption{Resistivity of 4 samples versus temperature in log scale
after subtraction of electron-electron and electron-phonon
contributions. Dashed line: Fit to (2) for sample 1,
$\rho_{K}=7.0$n$\Omega$cm and $T_{K}$=9.5K.} \label{Fig.6}
\end{figure}

The influence of Cr impurities on the proximity effect is shown in
Fig. 7. The dots represent experimentally measured reduced
amplitude of magnetoresistance oscillations at $T=0.28$K, similar
to that shown in Fig. 3 for Cu/Ni samples.

\begin{figure}
\centerline{\psfig{file=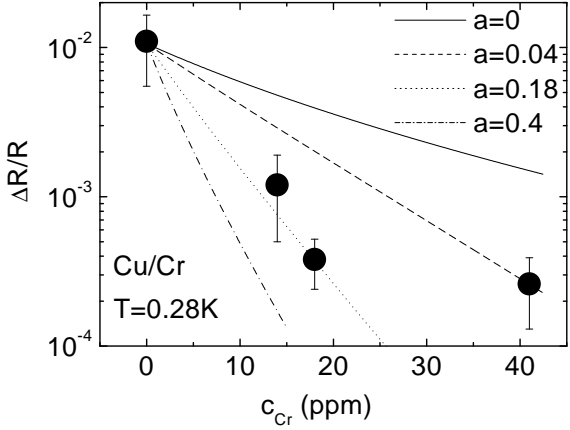,height=2.5 in}} \caption{Dots:
Reduced amplitude of resistance oscillations in logarithmic scale
as function of impurity concentration at $T=0.28K$ for Cu/Cr
samples. Lines: theoretical fits using $\L_{\phi 0}=1.6\mu$m and
parameter $a$=0 (solid), 0.04 (dashed), 0.18 (dotted), and 0.4
(dash-dotted).} \label{Fig.7}
\end{figure}

\section{CONDUCTANCE CALCULATION}
In order to calculate the correction to the conductance in our
structures due to the proximity effect we used Usadel's equation
for quasiclassical Green functions induced in the $N$ wire.
According to the theory, in the case of a weak superconducting
proximity effect, the amplitude of resistance oscillations can be
presented as (see e.g. \cite{21})
\begin {equation}
\frac{\Delta
R(V,T)}{R_{N}}=\frac{2\int_{0}^{\infty}F(\epsilon,V,T)m(\epsilon)d\epsilon}{eV},
\end{equation}

where 2$V$ is the total voltage applied between the two $N$
reservoirs, $\epsilon$ is the quasiparticle energy, $F(V,T)$ is
the difference of equilibrium distribution functions in $N$
reservoirs
\begin{equation}
F(V,T)=\frac{tanh(\frac{\epsilon+eV}{2k_{B}T})-tanh(\frac{\epsilon-eV}{2k_{B}T})}{2},
\end{equation}

and energy-dependent function
$m(\epsilon)=(1/16)(m_{1}(\epsilon)+m_{2}(\epsilon)+m_{3}(\epsilon)+m_{4}(\epsilon))$
determines averaged over the length $(L_{1}, L)$ correction to the
conductance due to proximity effect. Here 2$L$ is the distance
between normal reservoirs, 2$L_{1}$ is the distance between $N/S$
contacts, and $L_{2} = L - L_{1}$.

\begin{widetext}
\begin {equation}
m_{1}(\epsilon)=Re\left[\frac{F_{0}^{2}}{\theta^{2}}\frac{sinh2\theta_{1}sinh^{2}
\theta_{2}+sinh2\theta_{2}sinh^{2}
\theta_{1}-2\theta_{1}sinh^{2}\theta_{2}-2\theta_{2}sinh^{2}\theta_{1}}{2\theta
sinh^{2}{\theta}}\right],
\end{equation}

\begin {equation}
m_{2}(\epsilon)=Re\left[\frac{F_{0}^{2}}{\theta^{2}}\frac{2\theta_{2}cosh^{2}
\theta_{1}-sinh2\theta_{2}cosh^{2}
\theta_{1}-2\theta_{1}sinh^{2}\theta_{2}-sinh2\theta_{1}sinh^{2}\theta_{2}}{2\theta
cosh^{2}{\theta}}\right],
\end{equation}

\begin {equation}
m_{3}(\epsilon)=\left|\frac{F_{0}}{\theta}\right|^{2}
\left[\left|\frac{sinh\theta_{2}}{sinh\theta}\right|^{2}
\left(\frac{sinh2\theta^{'}_{1}}{2\theta^{'}}-\frac{sin2\theta^{''}_{1}}{2\theta^{''}}\right)+
\left|\frac{sinh\theta_{1}}{sinh\theta}\right|^{2}
\left(\frac{sinh2\theta^{'}_{2}}{2\theta^{'}}-\frac{sin2\theta^{''}_{2}}{2\theta^{''}}\right)\right],
\end{equation}

\begin {equation}
m_{4}(\epsilon)=\left|\frac{F_{0}}{\theta}\right|^{2}
\left[-\left|\frac{sinh\theta_{2}}{cosh\theta}\right|^{2}
\left(\frac{sinh2\theta^{'}_{1}}{2\theta^{'}}+\frac{sin2\theta^{''}_{1}}{2\theta^{''}}\right)-
\left|\frac{cosh\theta_{1}}{cosh\theta}\right|^{2}
\left(\frac{sinh2\theta^{'}_{2}}{2\theta^{'}}-\frac{sin2\theta^{''}_{2}}{2\theta^{''}}\right)\right],
\end{equation}

\end{widetext}
where $F_{0}^{2}=\Delta^{2}/[\Delta^{2}-(\epsilon+i\Gamma)^{2}]$
is the equilibrium condensate functions in the superconductor,
$\Gamma$ is the depairing rate in the superconductor. Here we
neglect the attenuation of condensate functions at the interface
because in all our samples $R_{N}$ was more than 10 times bigger
than interface resistance, $R_{int}$. Energy dependence of
condensate functions is determined by parameters $\theta$ and
$\theta_{1,2}$,

\begin{equation}
\theta=\theta^{'}+i\theta^{''}=\sqrt{\frac{2i\epsilon}{E_{Th}}+\left(\frac{L}{L_{\phi}}\right)^{2}},
\end{equation}

$\theta_{1,2}=\theta L_{1,2}/L$, $L_{\phi}=(D\tau_{\phi})^{1/2}$
is the phase breaking length in the normal wire and $E_{Th}=\hbar
D/L^{2}$ is the Thouless energy.

The effect of impurity of concentration can be included in the
model by reducing $L_{\phi}$ as follows:

\begin{equation}
\left(\frac{L}{L_{\phi}c}\right)^{2}=\left(\frac{L}{L_{\phi 0
}}\right)^{2}(1+bc)(1+ac),
\end{equation}

where $L_{\phi 0}$ is $L_{\phi}$ for pure Cu and $L_{\phi c}$ is
$L_{\phi}$ at impurity concentration $c$. Constants $b$ and $a$
describe change in elastic scattering rate and dephasing rate
respectively, according to the following linear approximations:

\begin{equation}
\tau^{-1}_{ec}=\tau^{-1}_{e0}(1+bc),
\end{equation}

\begin{equation}
\tau^{-1}_{\phi 0c}=\tau^{-1}_{\phi 0}(1+ac),
\end{equation}

where again subscripts "0" and "c" correspond to values for pure
Cu and for impurity concentration $c$. The value of $b$ can be
obtained from Fig. 4. Using experimental values for sample
parameters and $L_{\phi 0}$ and $a$ as fitting parameters the
experimental dependence of proximity effect on impurity
concentration can be compared to the one predicted by theory. The
value of $a$ is only constant in the single impurity regime. The
impurity-impurity interaction leads to the dependence $a=a(c)$. In
this way the spin-glass transition can be included in the model as
well. This effect is seen for 42 ppm sample in Fig. 5, where the
curve obtained in linear approximation does not go through this
point.

\section{ANALYSIS AND DISCUSSION}

Lines in Fig. 3 show results of fitting of experimental data for
Cu/Ni samples by the theory as described in section IV. The best
fit was obtained for $L_{\phi 0}=1.9\mu$m and $a$=35. This
corresponds to $\tau^{-1}_{\phi 0}=3.0\times10^{9}$s$^{-1}$. The
extra dephasing in Cu/Ni sample with 4$\%$Ni is
$\Delta\tau^{-1}_{\phi}=4.2\times10^{9}$s$^{-1}$. The same
calculation for Cu/Cr samples is shown in Fig. 6. The best fit was
obtained for $L_{\phi 0}=1.6\mu$m and $a$=0.18. This corresponds
to $\tau^{-1}_{\phi 0}=7\times10^{9}$s$^{-1}$. The extra dephasing
in Cu/Cr sample with 14 ppm of Cr is
$\Delta\tau^{-1}_{\phi}=1.7\times10^{10}$s$^{-1}$. Comparing the
above two samples one can see that the dephasing in 14ppm Cr
samples is about 4 times higher than that in 4$\%$ Ni sample, and
the Kondo slope the former is about 3.5 times higher than that in
the former (see Fig. 5). This suggests that the dephasing in Cu/Ni
samples can be attributed to the small amount of Cr impurities
introduced during Ni deposition. Note that pure Cu samples do not
show the Kondo effect within our experimental accuracy, which
means that the amount of Cr impurities can be proportional to Ni
concentration explaining the dependence of dephasing on
concentration of Ni. In our analysis of sources of dephasing we
can neglect contributions from electron-phonon interaction which
is of the order of 10$^{6}$s$^{-1}$ (Ref. 22) and due to
electron-electron interaction which is of the order of
10$^{8}$s$^{-1}$ (Ref. [23]) at the temperature of our experiment.
Therefore, we assume that additional dephasing is entirely due to
magnetic impurities.

For Cu/Ni system impurity spin $S$=1/2 (Ref. [24]) so that the
spin flip rate in the limit $T \ll T_{K}$ can be calculated using
the following expressions \cite{12}

\begin {equation}
\tau_{sf}^{-1}=\frac{9\pi^{3}}{8}\frac{\epsilon_{F}}{\hbar}\frac{T^{2}}{T_{K}^{2}}\times
c,
\end{equation}

where $\epsilon_{F}=1.1\times 10^{-18}J$ is the Fermi energy of Cu
\cite{18}. For 4$\%$ Ni sample (13) gives
$\tau_{sf}^{-1}$=1.4$\times10^{9}$s$^{-1}$. This is smaller than
the dephasing rate due to Cr impurities.

For Cu/Cr system the spin flip rate at $T < T_{K}$ can be written
as follows \cite{11}

\begin {equation}
\tau_{sf}^{-1}=8\pi\frac{\epsilon_{F}}{\hbar}\frac{S^{2}-1/4}{ln^{2}(T_{K}/T)}\times
c,
\end{equation}

where $S=3/2$ is Cr impurity spin \cite{20}. The dephasing rate
due to spin flip scattering can be found as \cite{11}

\begin {equation}
\tau_{\phi}^{-1}=\tau_{sf}^{-1}\frac{2}{\left(1+\frac{\alpha}{2}\frac{T_{SG}}{T}\right)^{2}},
\end{equation}

where $\alpha=1.85$ for $S=3/2$ and denominator in right-hand side
of (15) accounts for the reduction of dephasing due to the
spin-glass transition at the temperature $T_{SG}$. Substituting
$\tau_{\phi}^{-1}=1.7\times10^{10}$s$^{-1}$ for Cu/Cr sample with
$c=14\times10^{-6}$ at $T$ = 0.3 K into (14) and (15) one gets
$T_{SG}$ to be about 2K in reasonable agreement with resistivity
measurements shown in Fig. 5.

In conclusion, we have studied the dephasing of conduction
electrons by magnetic impurities in Cu/Ni and Cu/Cr samples.
Dephasing in Cu/Cr samples is associated with spin-flip scattering
due to the Kondo effect. Influence of spin glass transition on the
superconducting proximity effect has been observed. Estimation of
$T_{SG}$ made using formulas for spin-flip rate and the dephasing
rate are in agreement with experiment. The dephasing in Cu/Ni
samples can be explained by the contribution of other magnetic
impurities, such as Cr by comparison to results in Cu/Cr samples.

\section{ACKNOWLEDGMENTS}

We thank L.I. Glazman for valuable discussions. The work was
supported by EPSRC grant AF/001343. A.F.V. thanks the DFG for the
financial support within the SFB 491 and Mercator-Gastprofessoren.

\end{document}